# On-chip optical wavefront shaping by transverse spin induced Pancharatanam-Berry phase

Wanyue Xiao[1] and Shubo Wang[1,2,*]

[1]Department of Physics, City University of Hong Kong, Tat Chee Avenue, Kowloon, Hong Kong, China
[2]City University of Hong Kong Shenzhen Research Institute, Shenzhen, Guangdong 518057, China

*Corresponding author: shubwang@cityu.edu.hk

**Abstract**

Pancharatnam-Berry (PB) metasurfaces can be applied to manipulate the phase and polarization of light within subwavelength thickness. The underlying mechanism is attributed to the geometric phase originating from the longitudinal spin of light. Here, we demonstrate a new type of PB geometric phase derived from the intrinsic transverse spin of guided light. Using full-wave numerical simulations, we show that the rotation of a metallic nano bar sitting on a metal substrate can induce a geometric phase covering $2\pi$ full range for the surface plasmons carrying intrinsic transverse spin. Specially, the geometric phase is different for the surface plasmons propagating in opposite directions due to spin-momentum locking. We apply the geometric phase to design metasurfaces to manipulate the wavefront of surface plasmons to achieve steering and focusing. Our work provides a new mechanism for on-chip light manipulations with potential applications in designing ultra-compact optical devices for imaging and sensing.

Manipulating the phase of electromagnetic waves in conventional optical structures strongly relies on the refractive indices of materials and the geometric dimensions of the structures. This usually results in bulky optical structures that face difficulties in miniaturizing and integrating into compact on-chip devices. A novel solution extensively explored in recent years is the so-called metasurfaces, which are two-dimensional periodic structures consisting of subwavelength dielectric or metallic units (i.e., meta-atoms) [1-3]. Metasurfaces can enable almost arbitrary manipulation of wavefront with various phases, including resonance phase [4-6], propagation phase [7-9], and Pancharatnam-Berry (PB) phase [10-14]. The PB phase is particularly attractive since it can cover $2\pi$ full range within a broadband of frequencies. The PB phase is a kind of geometric phase emerging in polarization evolutions on the Poincaré sphere [12, 15, 16]. It can be induced when circular-polarized incident light, which carries longitudinal spin parallel or anti-parallel to the propagating direction, passes through an anisotropic material or structures with spatially varying principle axis, for example, in metasurfaces composed of meta-atoms with spatially varying orientation angle [10, 11, 13]. This longitudinal spin induced PB phase is generally twice of the orientation angle of the principal axis. Such PB metasurfaces can give rise to counterintuitive phenomena such as anomalous reflection/refraction [10, 11] of light and remarkable devices such as flat metalenses for light focusing and imaging [8, 14, 17].

Conventional metasurfaces are designed for manipulating free-space propagating waves. Recently, increasing attention has been paid to on-chip metasurfaces for controlling guided waves, which have essential applications in photonic integrated circuits for optical computations and communications as well as quantum information processing [18-22]. Due to the less degrees of freedom in guided waves with constraints in at least one spatial dimension, it is relatively difficult to manipulate the wavefront of guided waves. Nevertheless, several types of metasurfaces have been proposed for on-chip light bending and focusing, which employ gradient refractive index to control the propagation phase or gradient geometry to control the resonance phase [18-20]. Interestingly, PB phase can also be applied to manipulate guided waves [21, 22], which is attributed to the transverse spin arising from the interference of guided waves. This PB phase is inherently nonuniform in space and, like the PB phase induced by longitudinal spin, special mechanisms are required to separate the cross- and co-polarized fields, introducing further complexity to the systems. Notably, guided waves contain evanescent fields that are elliptically polarized and carry intrinsic transverse spin with an intriguing property of spin-momentum locking, i.e., the guided waves with opposite transverse spin propagate in opposite directions [23-25]. A variety of interesting phenomena associated with transverse spin have been discovered, including anomalous lateral optical forces [26, 27], quantum spin-Hall effect of light [28, 29], and arbitrary order non-Hermitian exceptional points [30, 31]. An interesting question is: can the transverse spin give rise to PB phase for manipulating the wavefront of guided waves?

In this Letter, we demonstrate that the transverse spin of evanescent fields indeed can give rise to the PB geometric phase, and we propose a metasurface composed of simple metal nano bars for shaping the wavefront of surface plasmons based on the transverse spin induced PB phase. We show that the rotation of the nano bars can enable direct control of the PB phase covering full range of $2\pi$. The metasurface can be applied to realize on-chip bending and focusing of the SPPs.

We consider a bare plasmonic gold substrate immersed in the air under the illumination of a linearly polarized plane wave propagating in *yz*-plane. We set the wavelength $\lambda_0$ =633 nm, at which the gold has a relative permittivity $\varepsilon = -12$ (we neglect loss for simplicity) [32], and conduct numerical simulations using a finite element package COMSOL. Due to the interference of the incident and reflected fields, the total field has a spatially varying polarization that can be controlled by varying the $E_\perp$ and $E_{//}$ components (perpendicular and parallel to the incident plane, respectively) of the incident electric field. For $E_{//}/E_\perp = 2.1$, $\theta_i = 30°$ and at the height $h$ = 60 nm above the substrate, we have $|E_z/E_x| = 1$ and $\arg(E_z/E_x) = 90°$, i.e., the total field is circularly polarized in *xz*-plane, as denoted by the green circle in Fig. 1(a).

We arrange subwavelength metallic nano-bars on the substrate periodically along *y* direction with a period of *p*, as shown in Fig. 1(a). The nano-bars are made of gold with dimensions $l \times w \times t = 100 \text{ nm} \times 20 \text{ nm} \times 20 \text{ nm}$. The insets in Fig. 1(a) show the *xz* view and *yz* view of the unit cell. The centers of the nano-bars are all located at *h* = 60 nm, where the background field is circularly polarized. The orientation of the nano-bars in the *xz* plane is characterized by the angle $\alpha$ with respect to the horizontal direction. Such meta-atoms with different orientations may be fabricated by using 3D printing, such as the metal 3D nanoprinting technique in Ref. [33] with nanoscale precision. In the technique, two flows are used to create a double-layer system, comprising a top layer with the nanoparticles (NPs) and a bottom layer without the NPs that prevent NPs diffusion to the printed area. By controlling the applied electric field which serves as the "field maps" for the NPs and adjusting the flow field, the NPs of particular size can be precisely selected as building blocks to construct a variety of 3D nanostructures with complex geometry.

Under the excitation of the background field, the nano-bar unit gives rise to an electric dipole with two orthogonal components $p_l$ (in the direction parallel to the nano-bar) and $p_w$ (in the direction perpendicular to the nano-bar), which have different amplitudes and phases due to the anisotropy of the nano-bar, as shown in the inset in Fig. 1(b). The lines in Figure 1(b) show the relative amplitude and phase of $p_l$ and $p_w$ as a function of the rotation angle. We notice that $\text{abs}(p_l/p_w) \approx 6$ and $\arg(p_l/p_w) \approx -90$ degrees for $\alpha \in [0,180]$ degrees. Thus, the induced electric dipole moment **p** is elliptically polarized. Figure 1(c) shows the phases of $p_l$ and $p_w$ for different rotation angles.

Notably, they both change with the rotation angle $\alpha$ linearly, indicating that the phase of **p** has a simple linear dependence on $\alpha$. The fields of **p** will couple to the guided wave channels, i.e. the SPPs on the substrate surface, which has an effective index $n_s = \sqrt{\varepsilon/(\varepsilon+1)} = 1.04$ [23]. The SPPs are elliptically polarized and carry a transverse spin with the direction locked to their propagating direction, corresponding to the spin-momentum locking of transverse spin-orbit interactions [24, 28]. Thus, the two SPPs propagating in $+x$ and $-x$ directions excited by **p** carry opposite spins, as shown in Fig. 1(a) by the red and blue arrows [23, 24]. Notably, the SPP propagating in $+x$ direction and the background field have opposite spin directions. This spin flipping mediated by the nano-bars can give rise to a PB phase covering full range of 360°, as will be shown below.

The PB phase can be determined by measuring the phase of the SPPs for different rotation angle $\alpha$ of the nano-bar. The results are shown in Fig. 1(d) for the SPPs propagating in $+x$ (red solid line) and $-x$ (blue solid line) directions. For the SPP in $+x$ direction, the PB phase can cover 360°, while it is not the case for the SPP in $-x$ direction. In addition, we notice that the PB phase of the $+x$ propagating SPP slightly deviates from a linear relation with the rotation angle (i.e., $\varphi = 2\alpha$), which is different from the geometric phase in conventional PB metasurfaces. The reason is that the SPPs are elliptically polarized, while the transmitted light in conventional PB metasurfaces is circularly polarized.

The PB phase can be intuitively understood with the Poincaré polarization sphere [12, 15]. The Strokes vector for the SPPs is $\mathbf{S}^s = (S_1^s, S_2^s, S_3^s) = (\frac{-k_0^2}{k_x^2+k_z^2}, 0, \frac{\pm 2k_xk_z}{k_x^2+k_z^2})$, where '+' ('−') is for the SPP propagating in $-x$ ($+x$) direction; $k_0$ is the wavenumber in free space; $k_x$ and $k_z$ are wavevector components in the $x$ and $z$ directions satisfying $k_x^2 - k_z^2 = k_0^2$ [23]. The polarization of the induced electric dipole **p** can be characterized by a Strokes vector $\mathbf{S}^d = (S_1^d, S_2^d, S_3^d)$ with $S_1^d = \cos(2\alpha)\cos[2\tan^{-1}(|p_w/p_l|)]$, $S_2^d = \sin(2\alpha)\cos[2\tan^{-1}(|p_w/p_l|)]$, and $S_3^d = -\sin[2\tan^{-1}(|p_w/p_l|)]$, where $|p_w/p_l| = 1/6$ for the designed nano-bar here. Figures 2(a) and 2(b) show the evolution of the polarization states for the SPPs propagating in $+x$ and $-x$ directions, respectively. The polarization states of the background field, the electric dipole **p** and the SPPs are denoted by the points A, B and C on the Poincaré sphere, respectively. The PB geometric phase due to the polarization evolution A → B → C is determined by the solid angle subtended by the area enclosed by the loop A → B → C → A. Variation of the rotation angle (i.e., $\Delta\alpha$) of the nano-bar induces a change of its polarization state from point B to point B'. The change of the PB phase corresponds to half of the solid angle $\Omega$ formed by the closed loop A → B → C → B′ → A [12, 15, 16]. For the $+x$ propagating SPP, the maximum solid angle can reach $4\pi$, as shown by the inset in Fig. 2(a). In contrast, for the $-x$ propagating SPP, the maximum solid angle is less than $4\pi$, as shown by the inset in Fig. 2(b). We evaluate the solid angle to obtain the geometric phase and compare it to the phase directly obtained from the SPPs in the numerical simulations. The results are shown in Fig. 2(c) as the symbol and solid lines, which have good consistency.

We apply the transverse spin induced PB phase to design metasurfaces to manipulate the wavefront of the SPPs. The incident light is same as that in Fig. 1. We focus on the SPPs propagating in $+x$ direction since the modulation of its phase by the nano-bars can cover 360°. Figure 3(a) shows the $E_z$ field pattern induced by the metasurface composing of vertically oriented nano-bars, where the incident angle is $\theta_i = 30°$ and the excited SPPs propagate in a direction forming angle $\theta_r$ with the x-axis. The relationship between the incident angle $\theta_i$ and the deflection angle $\theta_r$ can be expressed as: $k_0 \sin\theta_i = k_{\text{spp}} \sin\theta_r$. Here, $k_0$ is the wavevector of the incident light; $k_{\text{spp}} = n_s k_0$ is the propagation constant of the SPPs. To achieve anomalous deflection, we consider a supercell with $m$ sub-units, as shown in Fig. 3(b). The period of the sub-units is $p = \lambda/3 = 202$ nm, where $\lambda$ is the wavelength of the SPPs. The rotation angle of the nano-bar $\alpha(y)$ is a function of $y$ coordinates of the nano-bars. The geometric phase difference between two nearby sub-units is $\Delta\varphi = \varphi[\alpha(y)] - \varphi[\alpha(y-p)]$, where $\varphi$ is the transverse spin induced PB phase. By arranging this supercell periodically along $y$ direction, we obtain a gradient metasurface. For a constant phase gradient along the $y$ direction introduced by the metasurface, we have the relationship $-k_0 \sin\theta_i + k_g = -k_{\text{spp}} \sin\theta_r$, where $k_g = \Delta\varphi / p$ is the phase gradient. Thus, deflection angle is $\theta_r = \sin^{-1}[(-k_g + k_0 \sin\theta_i)/k_{\text{spp}}]$. We design five metasurfaces with different phase gradient $k_g$ to demonstrate anomalous deflection of the SPPs. The simulated $E_z$ field of the SPPs is shown in Figs. 3(c)-(g) for $k_g = 2\pi/3p, 2\pi/6p, -2\pi/9p, -2\pi/6p$ and $-2\pi/5p$, respectively. The deflection direction predicted by the above analytical expression of $\theta_r$ is indicated by the yellow arrows, which agree with the simulated wavefronts. Specifically, when the phase gradient $k_g$ fulfills $-k_0 \sin\theta_i + k_g = -k_{\text{spp}}$, the SPPs will propagate along the metasurface (in the -y direction), corresponding to Fig. 3(f). Moreover, when $k_g < k_0 \sin\theta_i - k_{\text{spp}}$ or $k_g > k_0 \sin\theta_i + k_{\text{spp}}$, the SPP propagating in the $+x$ half plane cannot be excited, as shown in Fig. 3(g) for the case of $k_g < k_0 \sin\theta_i - k_{\text{spp}}$. We note that $k_g$ has different values for the SPPs propagating in the $+x$ and $-x$ half planes due to their different PB phases. Consequently, for a fixed incident angle, it is possible to excite unidirectional SPP propagating in $+x$ or $-x$ half plane only.

The transverse spin induced PB phase can be applied to realize on-chip metalens that can convert the incident plane wave to the SPPs converge at a desired focal point. The required phase profile at each location of the nano-bar array should satisfy $\varphi(y) = -k_0(\sqrt{y^2 + f^2} - f)$ [17] where $f$ is the working focal length of the metalens, and $y$ represents the location of the nano-bars. For oblique incident light with incident angle $\theta_i$, one needs to compensate the phase gradient introduced by the incident wavevector, and the required phase profile becomes $\varphi(y) = -k_0(\sqrt{y^2 + f^2} - f) + k_0 \sin\theta_i y$. The compensation can be provided by the transverse spin induced PB phase via the rotation of the nano-bars. Figure 4(a) shows the designed metalens, which includes 17 nano-bars with a sub-unit period of $p = \lambda/3$. Figure 4(b) shows the simulated $E_z$ field of the SPPs in $xy$ plane. We observe a focusing of the SPPs with a focal length of $5\lambda$. In addition, Fig. 4(c) plots the normalized intensity along the $y$ axis at the focal plane [indicated by the red dash line in Fig. 4(b)]. The results demonstrate that on-chip focusing of SPPs can be achieved with the designed metalens. We note that the above wave manipulations (including wave deflection) can also be achieved with the metasurfaces at other frequencies although the PB phase is generally different. Besides, dielectric meta-atoms and waveguides may also be used to achieve similar phenomena. Like conventional metasurfaces, the proposed metasurfaces have low efficiency, which can be improved by increasing the number of meta-atoms or exploiting their resonances.

In summary, we demonstrate that the transverse spin of light can give rise to the PB geometric phase. By using nano-bars sitting on a metal substrate under the illumination of a linearly polarized plane wave, we excite the SPPs propagating in two opposite directions, which carry different PB phases due to the spin-momentum locking of the SPPs. The phases of the SPPs can be controlled via varying the rotation angle of the nano-bars, which enables manipulating the wavefront of the SPPs. The transverse spin induced PB phase well agrees with the solid angle subtended by the evolution trajectory of the polarization states on the Poincaré sphere. For the SPPs that flip the spin direction of the background illumination field, the PB phase can cover $2\pi$ range. We further demonstrate the applications of the transverse spin induced PB phase by designing gradient metasurfaces and metalens to realize on-chip bending and focusing of the SPPs wavefronts. Our study provides a new type of PB phase for light manipulations and can facilitate the development of multi-functional meta devices for various on-chip applications.

**Acknowledgements**

We thank Prof. D. P. Tsai for helpful discussions. The work described in this paper was supported by the Research Grants Council of the Hong Kong Special Administrative Region, China (Projects Nos. CityU 11306019, CityU 11301820, and AoE/P-502/20) and National Natural Science Foundation of China (No. 11904306).

**REFERENCES**

1. C. Jung, G. Kim, M. Jeong, J. Jang, Z. Dong, T. Badloe, J. K. W. Yang, and J. Rho, Chem. Rev. 121, 13013 (2021).
2. Q. Ma, G. D. Bai, H. B. Jing, C. Yang, L. Li, and T. J. Cui, Light Sci. Appl. 8, 98 (2019).
3. G. Li, S. Zhang, and T. Zentgraf, Nat. Rev. Mater. 2, 17010 (2017).
4. N. Yu, P. Genevet, M. A. Kats, F. Aieta, J. P. Tetienne, F. Capasso, and Z. Gaburro, Science 334, 333 (2011).
5. M. Decker, I. Staude, M. Falkner, J. Dominguez, D. N. Neshev, I. Brener, T. Pertsch, and Y. S. Kivshar, Adv. Opt. Mater. 3, 813 (2015).
6. E. Almeida, G. Shalem, and Y. Prior, Nat. Commun. 7, 10367 (2016).
7. M. Khorasaninejad, Z. Shi, A. Y. Zhu, W. T. Chen, V. Sanjeev, A. Zaidi, and F. Capasso, Nano Lett. 17, 1819 (2017).
8. B. H. Chen, P. C. Wu, V. C. Su, Y. C. Lai, C. H. Chu, I. C. Lee, J. W. Chen, Y. H. Chen, Y. C. Lan, C. H. Kuan, and D. P. Tsai, Nano Lett. 17, 6345 (2017).
9. S. Shrestha, A. C. Overvig, M. Lu, A. Stein, and N. Yu, Light Sci. Appl. 7, 85 (2018).
10. Z. e. Bomzon, G. Biener, V. Kleiner, and E. Hasman, Opt. Lett. 27, 1141 (2002).
11. M. Kang, T. Feng, H. T. Wang, and J. Li, Opt. Express 20, 15882 (2012).
12. E. Cohen, H. Larocque, F. Bouchard, F. Nejadsattari, Y. Gefen, and E. Karimi, Nat. Rev. Phys. 1, 437 (2019).
13. X. Wu, H. Cao, Z. Meng, and Z. Sun, Opt. Express 30, 15158 (2022).
14. I. Kim, J. Jang, G. Kim, J. Lee, T. Badloe, J. Mun, and J. Rho, Nat. Commun. 12, 3614 (2021).
15. M. V. Berry, J. Mod. Opt. 34, 1401 (1987).

16. S. Pancharatnam, Proc. Indian Acad. Sci. 44, 247 (1956).
17. S. Wang, P. C. Wu, V. C. Su, Y. C. Lai, M. K. Chen, H. Y. Kuo, B. H. Chen, Y. H. Chen, T. T. Huang, J. H. Wang, R. M. Lin, C. H. Kuan, T. Li, Z. Wang, S. Zhu, and D. P. Tsai, Nat. Nanotechnol. 13, 227 (2018).
18. Z. Wang, T. Li, A. Soman, D. Mao, T. Kananen, and T. Gu, Nat. Commun. 10, 3547 (2019).
19. R. Yang, Y. Shi, C. Dai, C. Wan, S. Wan, and Z. Li, Opt. Lett. 45, 5640 (2020).
20. Y. Fan, X. Le Roux, A. Korovin, A. Lupu, and A. de Lustrac, ACS Nano 11, 4599 (2017).
21. J. Ji, Z. Wang, J. Sun, C. Chen, X. Li, B. Fang, S. N. Zhu, and T. Li, Nano Lett. 23, 2750 (2023).
22. L. Li, T. Li, X. M. Tang, S. M. Wang, Q. J. Wang, and S. N. Zhu, Light: Science & Applications 4, e330-e330 (2015).
23. K. Y. Bliokh and F. Nori, Phys. Rev. A 85, 061801 (2012).
24. K. Y. Bliokh and F. Nori, Phys. Rep. 592, 1 (2015).
25. P. Shi, L. Du, C. Li, A. V. Zayats, and X. Yuan, Proc. Natl. Acad. Sci. U.S.A. 118, e2018816118 (2021).
26. S. B. Wang and C. T. Chan, Nat. Commun. 5, 3307 (2014).
27. F. Kalhor, T. Thundat, and Z. Jacob, Appl. Phys. Lett. 108, 061102 (2016).
28. K. Y. Bliokh, D. Smirnova, and F. Nori, Science 348, 1448 (2015).
29. B. Xie, G. Su, H. F. Wang, F. Liu, L. Hu, S. Y. Yu, P. Zhan, M. H. Lu, Z. Wang, and Y. F. Chen, Nat. Commun. 11, 3768 (2020).
30. S. Wang, B. Hou, W. Lu, Y. Chen, Z. Q. Zhang, and C. T. Chan, Nat. Commun. 10, 832 (2019).
31. H. Shi, Z. Yang, C. Zhang, Y. Cheng, Y. Chen, and S. Wang, Opt. Express 29, 29720 (2021).
32. J. Lin, J. P. B. Mueller, Q. Wang, G. Yuan, N. Antoniou, X. C. Yuan, and F. Capasso, Science 340, 331 (2013).
33. B. Liu, S. Liu, V. Devaraj, Y. Yin, Y. Zhang, J. Ai, Y. Han, and J. Feng, Nat. Commun. 14, 4920 (2023).

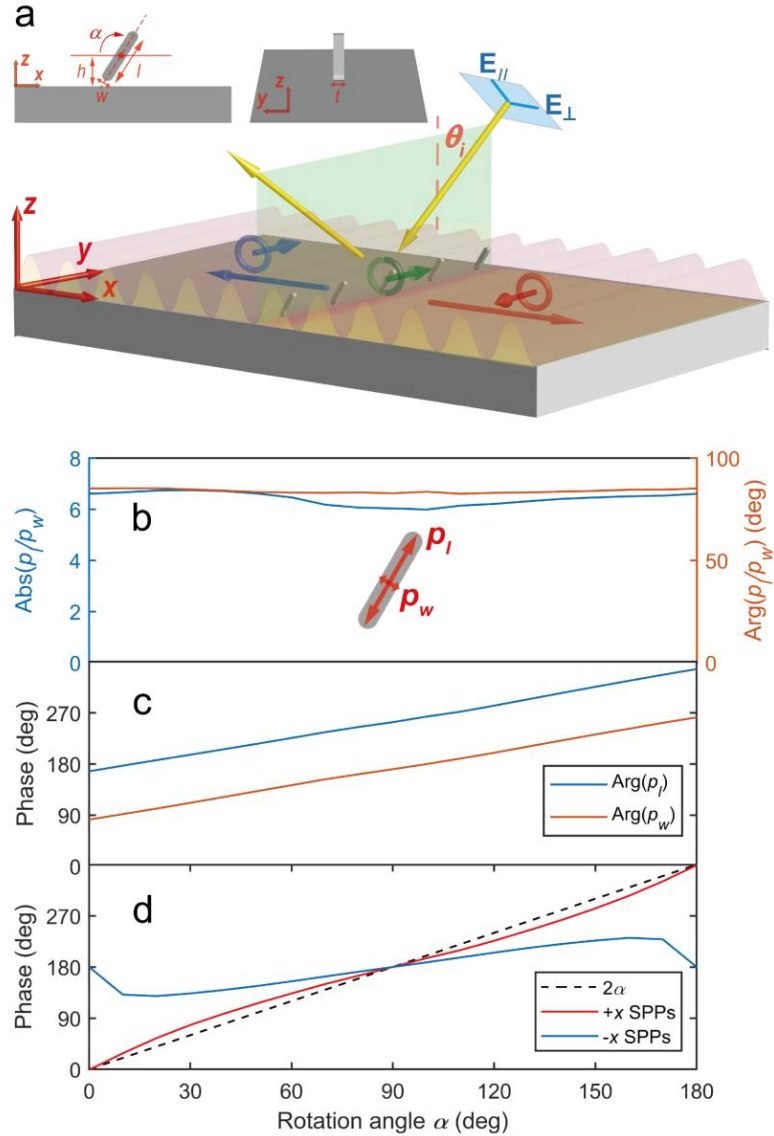

**Fig. 1.** (a) Schematic of the metasurface with transverse spin induced PB phase. The two insets show the dimensions and orientation of the metallic nano-bar. Here, $l = 110$ nm, $w = 20$ nm, $h = 60$ nm, and $t = 20$ nm. The nano-bars are periodically arranged on the substrate along $y$ axis with a period $p = \lambda/3$ ($\lambda = \lambda_0/n_s = 606$ nm is the wavelength of the SPPs). The incident light has $E_{//}/E_\perp = 2.1$ and $\theta_i = 30°$. The green arrow indicates the spin of the background field. The red and blue arrows denote the spins of the excited SPPs propagating in $+x$ and $-x$ direction, respectively. (b) Relative amplitude and phase of the induced electric dipole components $p_l$ and $p_w$ in the nano-bar. (c) The phases of $p_l$ and $p_w$ for different rotation angles. (d) The PB geometric phases of the SPPs propagating in $+x$ and $-x$ direction as a function of the rotation angle, which deviate from the conventional linear relation denoted by the dashed line.

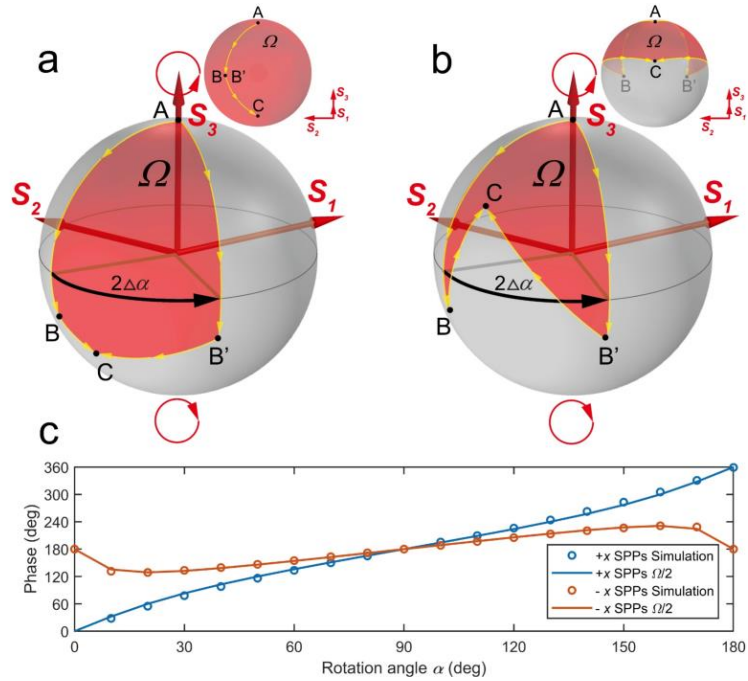

**Fig. 2.** The evolution trajectories of the polarization states on the Poincaré sphere for the SPPs propagating in (a) $+x$ and (b) $-x$ direction, respectively. The polarization states of the background field, the dipole **p** and the SPPs are marked by points A, B and C, respectively. A rotation of the nano-bar by angle $\Delta\alpha$ changes the polarization state from B to B′. The geometric phase equals to half of the solid angle defined by the closed loop A → B → C → B′ → A. The insets in (a) and (b) show the maximum solid angle $\Omega$ that can be achieved in each case. For (a), the maximum solid angle can reach $4\pi$. For (b), the maximum solid angle is less than $4\pi$. (c) Comparison between the PB phase obtained by evaluating the solid angle $\Omega$ and by the simulation.

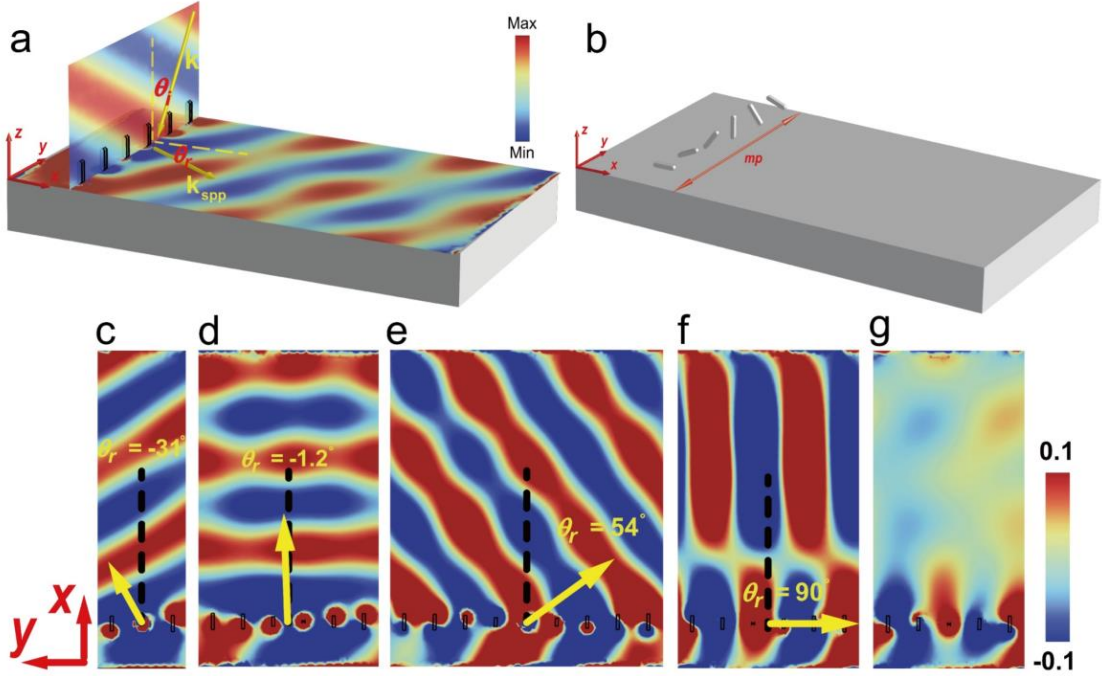

**Fig. 3.** (a) Deflection of the SPPs by the metasurface with zero phase gradient. (b) The supercell of the gradient metasurface. It is composed of $m$ nano-bars separated by $p = \lambda/3$. (c)-(g) show the bending of SPPs with different deflection angle $\theta_r$, which is achieved with metasurfaces with different phase gradient $k_g = \Delta\varphi / p$ and $\Delta\varphi$ is the geometric phase difference between two nearby sub-units. The field values are normalized by the $E_z$ component of the incident wave. (c) $m = 3, \Delta\varphi = 2\pi/3, k_g = 2\pi/3p$, and the deflection angle is $\theta_r = -34°$; (d) $m = 6, \Delta\varphi = 2\pi/6$, $k_g = 2\pi/6p$, and the deflection angle is $\theta_r = -1.2°$; (e) $m = 9, \Delta\varphi = -2\pi/9, k_g = -2\pi/9p$, and the deflection angle is $\theta_r = 54°$; (f) $m = 6, \Delta\varphi = -2\pi/6, k_g = -2\pi/6p$, and the SPPs propagate along -y direction; (g) $m = 5$, $\Delta\varphi = -2\pi/5, k_g = -2\pi/5p$, and the SPPs cannot propagate in the +$x$ half plane.

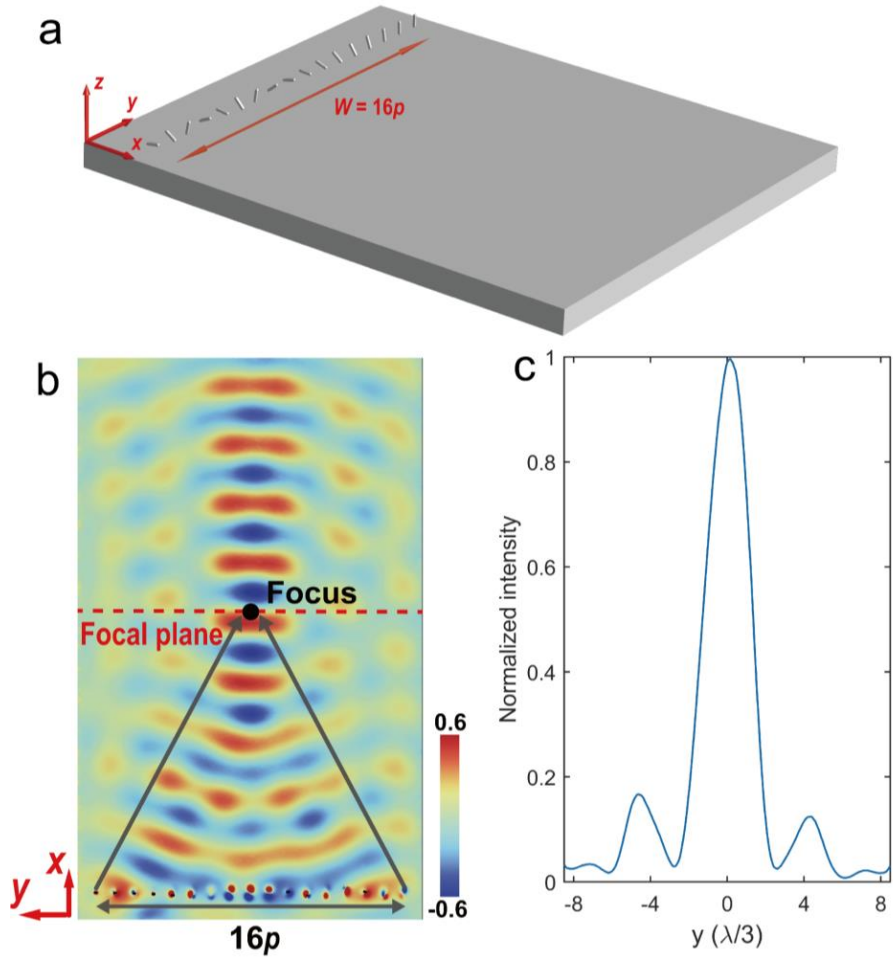

**Fig. 4.** (a) Schematic of the on-chip metalens. The metalens includes 17 nano-bars with a sub-unit period $p = \lambda/3$. (b) The $E_z$ field of the SPPs in the $xy$ plane. The red dash line marks the focal plane. The field values are normalized by the $E_z$ component of the incident wave. (c) The normalized intensity of SPPs along the $y$ axis at the focal plane.